\documentclass[aps,prd,10pt,twocolumn,superscriptaddress,nobibnotes,longbibliography,nofootinbib]{revtex4-2}
\pdfoutput=1

\usepackage[nodayofweek]{datetime} 

\newdateformat{daymonthyeardate}{\THEDAY\;\monthname[\THEMONTH], \THEYEAR}
\usepackage{mathtools}
\usepackage{dcolumn}% Align table columns on decimal point
\usepackage{bm}% bold math
\usepackage{amsmath}
\usepackage{amssymb}
\usepackage{graphicx}

\usepackage[colorlinks=true, allcolors=blue]{hyperref}

\begin{document}

\title{Parameter Estimation for Stellar-Origin Black Hole Mergers In LISA}
\author{Matthew C. Digman}
\email{matthew.digman@montana.edu}

\affiliation{eXtreme Gravity Institute, Department of Physics, Montana State University, Bozeman, Montana 59717, USA}

\author{Neil J. Cornish}

\affiliation{eXtreme Gravity Institute, Department of Physics, Montana State University, Bozeman, Montana 59717, USA}
%\affiliation{Department of Physics, Ohio State University, Columbus, OH 43210}

\date{\daymonthyeardate\today}

\begin{abstract}
The population of stellar origin black hole binaries (SOBHBs) detected by existing ground-based gravitational wave detectors is an exciting target for the future space-based Laser Interferometer Space Antenna (LISA). LISA is sensitive to signals at significantly lower frequencies than ground-based detectors. SOBHB signals will thus be detected much earlier in their evolution, years to decades before they merge. The mergers will then occur in the frequency band covered by ground-based detectors. Observing SOBHBs years before merger can help distinguish between progenitor models for these systems. We present a new Bayesian parameter estimation algorithm for LISA observations of SOBHBs that uses a time-frequency (wavelet) based likelihood function. Our technique accelerates the analysis by several orders of magnitude compared to the standard frequency domain approach and allows for an efficient treatment of non-stationary noise.
\end{abstract}

\maketitle

%%%%%%%%%%%%%%%%%%%%%%%%%%%%%%%%%%%%%%%%%%%
%%%%%%%%%%%%%%%%%%%%%%%%%%%%%%%%%%%%%%%%%%%

\section{Introduction}\label{sec:intro}
The Laser Interferometer Space Antenna (LISA) will be the first space-based gravitational-wave observatory when it launches in the mid-2030s \cite{2017arXiv170200786A}. An important class of sources LISA will detect are stellar-origin black hole binaries (SOBHBs), observed during the early inspiral phase, years to decades before merger. These systems are from the same population currently being explored by existing ground-based detectors such as LIGO and Virgo, starting in 2015 with GW150914 \cite{LIGOScientific:2016aoc}. 

Ground-based gravitational-wave detectors have detected approximately 100 mergers of stellar mass binaries to date \cite{LIGOScientific:2021djp,Nitz:2021zwj,Olsen:2022pin,Venumadhav:2019lyq}. A subset of these binaries would also have been detectable several years in advance by LISA while their gravitational-wave frequencies remained in the mHz \cite{Sesana:2016ljz}. SOBHBs will likely represent the bulk of gravitational-wave sources jointly detectable by both ground and space-based instruments, though  other source classes such as intermediate-mass black hole binaries \cite{Amaro-Seoane:2009vjl,Jani:2019ffg,Sedda:2021yhn,Saini:2022hrs} and bursts from cosmic strings \cite{ShapiroKey:2008ckh} are also possible multi-wavelength targets. 

The prospect of multi-wavelength gravitational-wave observations makes possible several opportunities for gravitational-wave science. Multi-wavelength observations open opportunities for robust tests of general relativity (GR) \cite{Toubiana:2020vtf,Gnocchi:2019jzp,Carson:2019rda,Vitale:2016rfr,Barausse:2016eii,Chamberlain:2017fjl,LISA:2022kgy,Nakano:2021bbw,Berti:2004bd}. For example, in many cases, LISA data will forecast the merger time to within a few seconds months or even years in advance of an actual SOBHB merger. If energy emission channels beyond those predicted by GR exist, they would shift the observed time of merger relative to the GR forecast. If such deviations are consistent with environmental factors, such multi-wavelength observations could provide compelling motivation for targeting multi-messenger searches for electromagnetic counterparts \cite{Caputo:2020irr,Barausse:2014tra,Barausse:2014pra,Tamanini:2019usx,Cardoso:2019rou}. The combined localizations and parameter estimation for multi-wavelength sources will typically be drastically better than is presently achieved with ground-based instruments alone \cite{Liu:2020eko,Liu:2020nwz,Sathyaprakash:2019rom,Cutler:1997ta}. The improved localizations will permit narrow-field electromagnetic telescopes to make far deeper observations to have a higher chance of detecting or tightly constraining any potential electromagnetic counterparts to SOBHB mergers. It will also be possible to detect non-zero eccentricity and spin precession, which will place constraints on binary environment and formation scenarios \cite{Franciolini:2021xbq,Klein:2022rbf,Postnov:2014tza,Benacquista:2011kv,Su:2021dwz,Bird:2016dcv,Antonini:2012ad,Samsing:2017xmd,Gerosa:2018wbw,Gerosa:2019dbe}. 

The combined constraints can also be used for black-hole spectroscopy to detect or constrain possible non-GR effects in the post-merger ringdown \cite{CalderonBustillo:2020rmh,Bhagwat:2021kwv,Berti:2005ys}. With months-in-advance merger alerts from LISA, future-generation ground-based detectors may be able to alter their detector configuration in expectation of a merger to better optimize the detector for black hole spectroscopy of a particular source \cite{Tso:2018pdv,Srivastava:2019fcb,PhysRevD.103.124043}. LISA pre-merger alerts can also help inform scheduled ground-based detector downtimes to avoid missing multi-wavelength science opportunities. 

Parameter estimation for SOBHBs in LISA presents a unique data analysis challenge. SOBHB sources in the mHz band will evolve significantly over several years, so the simple waveform models that are adequate for slow-evolving sources like galactic white dwarf binaries will be inadequate for SOBHBs. However, SOBHBs also evolve too slowly to ignore the variation in the detector response function over the lifetime of the signal or rotation of the constellation can, as is currently the case for near-merger SOBHBs in the band of existing ground-based detectors. While the rotation of the constellation is a powerful tool for constraining the location and polarization of SOBHB sources, it also means that the waveform model is no longer well-described by the long wavelength approximation, and reduces the inherent advantages of Fourier-domain-based searches. As an additional complication, SOBHBs will evolve through the cavity pole frequency, above which the detector response is complex \cite{Vecchio:2004vt,Vecchio:2004ec,Marsat:2018oam}.

In the presence of these constraints, it is necessary to develop techniques to evaluate likelihoods with high computational efficiency. One way to do this is by representing the data in a fashion that achieves some inherent signal compression into fewer real parameters. Data analysis techniques also need to be able to straightforwardly account for non-stationary noise. This paper describes a wavelet-domain-based parameter estimation pipeline that achieves both goals. The pipeline can efficiently generate large numbers of effective samples for a various sources in a tractable amount of compute time for LISA analysis. 

Nearby neutron star-neutron star and neutron star-black hole binaries are also interesting targets for multi-wavelength observation. In this work, we showcase the use of our parameter estimation pipeline for SOBHBs. Our pipeline should also perform well for lighter types of binaries as long as an adequate waveform model is available. Although observation of non-zero eccentricities in such systems is potentially useful for population studies, we leave accounting for eccentricity in LISA parameter estimation to future work. Throughout this work, we use a Python \cite{10.5555/1593511} implementation\footnote{Our Python IMRPhenomD implenetation is available at \url{https://github.com/XGI-MSU/PyIMRPhenomD}} of the IMRPhenomD 3.5PN \cite{Husa:2015iqa,Khan:2015jqa} waveform model, which does not include spin precession or eccentricity. The performance-sensitive portions of our Python code are accelerated using the Numba \cite{lam2015numba} just-in-time compiler.

LISA will likely experience a number of different kinds of non-stationary noise. Instrumental sources, including glitches, micrometeoroid impacts, off-gassing, thruster noise, repointing of the antenna for communications, aging of components, the solar system environment, and space weather, can all produce noise levels that vary with time  \cite{Edwards:2020tlp,Adams:2010vc,Adams:2013qma,Robson:2018jly,Armano:2016bkm,Armano:2018kix,LISAPathfinder:2019eny,Baghi:2019eqo,Purdue:2007zz,Littenberg:2010gf}. Astrophysical sources can also produce an unresolvable astrophysical time-varying gravitational-wave background. The rotation of LISA's antenna pattern relative to the galaxy will produce a gravitational-wave background that varies with time due to the high concentration of unresolvable white dwarf binaries towards the galactic center \cite{Cornish:2017vip,Digman:2022jmp,Timpano:2005gm,Karnesis:2021tsh,Cornish:2002bh,Cornish:2001hg,Cornish:2002bh,Breivik:2019lmt,Lamberts:2019nyk,Edlund:2005ye,Nissanke:2012eh,Renzini:2021iim}. Extragalactic sources \cite{Adams:2013qma,Boileau:2021sni,Bonetti:2020jku,2020MNRAS.491.4690M,Fan:2022wio} can produce a stochastic background that exhibits both intrinsic stochastic variation, as sources merge and fluctuate in amplitude, and periodic variation, as the LISA constellation rotations. 

Several previous analyses have considered the problem of Bayesian parameter estimation for LISA SOBHBs \cite{Toubiana:2022vpp,Toubiana:2020cqv,Buscicchio:2021dph,Marsat:2020rtl}. In this work, we present an efficient pure Python wavelet-domain parameter estimation pipeline, which is able to generate large numbers of effective samples in a relatively small amount of compute time. The wavelet domain gives our pipeline the flexibility to accept various non-stationary noise models as input. Our pipeline is also easily adaptable to handle a variety of other LISA sources. We use the rigid adiabatic approximation to the full LISA response, including the A, E, and T channels \cite{Rubbo:2003ap,Cornish:2020vtw}. 

The structure of the paper is as follows. In Sec.~\ref{sec:methods}, we describe our implementation of a wavelet domain parallel-tempered MCMC pipeline for parameter estimation. In Sec.~\ref{sec:results}, we show parameter estimation results for two different test sources: a source with parameters chosen to resemble GW150914 towards the end of its time in the LISA band; and a lower signal-to-noise (S/N) source in the presence of an aperiodic non-stationary noise background,  to demonstrate the utility of the wavelet domain in modeling non-stationarity. We also show some results for the computational efficiency of our pipeline. In Sec.~\ref{sec:conclusion}, we conclude and look to future work toward developing a full global fit pipeline for LISA data analysis.

\section{Methods}\label{sec:methods}

\subsection{Wavelet Domain Analysis}\label{ssec:wavelet}
Throughout this work, we use a Wilson-Daubechies-
Meyer (WDM) wavelet basis \cite{Necula:2012zz} with the normalization conventions used in \cite{2020PhRvD.102l4038C,Digman:2022jmp}. 

A key advantage of wavelet-based analysis for sources that undergo appreciable frequency evolution over the time of observation is that it inherently compresses the information as compared to either time or frequency domain analyses \cite{2020PhRvD.102l4038C}. 

For a concrete example of the compression inherent in the wavelet domain, consider a source with a frequency linearly increasing over range $\Delta F$ during an observation time $T_\text{obs}$ with sampling frequency $dt$. In the time domain, there is no compression, and all $N=T_\text{obs}/dt$ data samples are required to describe the source completely. In the frequency domain, the information about the source is compressed into a band of width $\Delta F$, such that the would be described by approximately $2T_\text{obs}\Delta F$ data samples in the Fourier Domain. As $\Delta F\longrightarrow 0$, the information is compressed into only a few data samples, and the Fourier representation becomes highly efficient. However, as the source chirps more, the Fourier representation becomes less efficient at compressing the data. In any event, the efficiency of both representations scales linearly with $T_\text{obs}$. 

If the source is instead represented on a WDM wavelet grid with $N_f\simeq\sqrt{T_\text{obs}/\Delta F}/(4 dt)$, $N_t\simeq 4\sqrt{\Delta F T_\text{obs}}$, the information about the source is compressed into a time-frequency track approximately two frequency pixels wide and $N_t$ time pixels long. The entire track is then compressed into approximately only  $8\sqrt{\Delta F T_\text{obs}}$ pixels. Therefore, in a well-chosen wavelet grid the number of parameters required to represent the source scales only $\propto \sqrt{T_\text{obs}}$. 

In this example, for a source evolving through a bandwidth of $\Delta F=10\;\text{mHz}$ over the course of a 2-year dataset sampled at $dt=1\,\text{s}$ intervals, the uncompressed time domain representation of the source would require $63,072,000$ data points. A frequency domain representation would compress the information by a factor of $\sim 50$, into $\sim 1,300,000$ data points. The particular choice of wavelet grid described above would require $\sim 6,400$ data points. Therefore the information in the wavelet domain would be represented approximately $200$ times as efficiently as in the frequency domain, and $10,000$ times as efficiently as in the time domain. If the number of operations required to perform likelihood calculations is reduced by a proportionate factor, a wavelet-domain-based parameter-estimation MCMC pipeline could be expected to be at least two orders of magnitude more efficient than either a time or frequency domain approach for a source like this. 

Evolving through $\mathcal{O}(10\;\text{mHz})$ in $\sim2$ years of data collection is fairly representative of a typical rate of evolution for anticipated LISA-detectable SOBHBs. Therefore, the calculation above is  useful as a ballpark estimate of the degree of compression that can be realistically expected for sources considered in this paper. 

Note that even for a nearly constant-frequency source where the Fourier domain representation would in principle be highly efficient, the noise over the course of the LISA mission will still be non-stationary, for example due to the time variation of the galactic stochastic background, as considered in \cite{Digman:2022jmp}. Such non-stationarities will induce a non-diagonal noise covariance matrix in the Fourier domain, which complicates analysis. However, provided the noise evolves adiabatically, choosing a grid resolution such that the noise evolves very little over the span of a single time pixel $dt N_f$ will diagonalize the noise covariance matrix in the wavelet domain \cite{2020PhRvD.102l4038C}, greatly simplifying likelihood calculations. Therefore, a wavelet-domain analysis is well-suited to most or all anticipated LISA source classes. 

Because different source classes will evolve at different rates, the optimal wavelet grid may depend somewhat on the source class, and even the individual source. Therefore an optimized global fit pipeline might involve a mix of different grid resolutions representing the same underlying data. However, for many applications, an approximately square grid of $N_f$ and $N_t$ is likely a suitable compromise.

\subsection{Parameterization}\label{ssec:parameterization}
There are several possible ways to specify the parameters to be searched over. Because we do not include spin precession or eccentricity, the system is described by 11 total parameters. The position on the sky is described by a distance, $D_l$, the ecliptic latitude $\cos\theta$ and the ecliptic longitude $\phi$. The orientation of the binary is described by an inclination, $\cos i$, the merger phase $\phi_c$, and a polarization angle $\psi$. 

For SOBHB sources in LISA, the initial frequency at the start of observations $F_i$ is more directly observable than the merger time. Therefore, it is more useful to conduct the MCMC search over $F_i$ than $t_c$, although because the IMRPhenomD waveform model predicts the relationship between $F_i$ and $t_c$ they can easily be interchanged for reporting of results. 

For the two spin parameters, there are a variety of possible choices. Searching directly over $\chi_1\in[-1,1]$ and $\chi_2\in[-1,1]$ is not ideal, because they are highly correlated. A better choice that removes some of the correlations could be $\chi_s=(\chi_1+\chi_2)/2$, $\chi_a=(\chi_1-\chi_2)/2$. However, there are still more natural ways to parameterize the waveform. In IMRPhenomD, the most natural spin parameters to describe the waveform are $\bar{\chi}_{\text{PN}}=\chi_s+\delta_m\chi_a/(1-76/113\eta)$ and $\chi_a$, where $\delta_m=\sqrt{1-4\eta^2}=(m_1-m_2)/M_t$ and $\eta=m_1 m_2/(m1+m2)^2/$ is the symmetric mass ratio. In this parametrization,  $\bar{\chi}_{\text{PN}}\in[-1,1]$, $\chi_a\in[-1,1]$. These are the two parameters we perform the MCMC search over internally. However, to more closely correspond to the results reported in existing LIGO/Virgo literature, we report our results in terms of the standard effective spin parameters, $\chi_\pm =(m_1\chi_1 \pm m_2\chi_2)/(m_1 + m_2)$. 

Lastly, we must choose the two mass parameters to search over. We search in terms of the total mass, $M_t=m_1+m_2$, and the chirp mass, $\mathcal{M}_c=(m_1m_2)^{3/5}/M_t^{1/5}$. For plotting purposes, we instead show results in terms of $\delta_m=(m_1-m_2)/M_t$ and $\mathcal{M}_c$, as for the sources plotted $\delta_m$ exhibits less dynamic range than $M_t$ and therefore makes the relevant probability contours easier to visualize.

\subsection{MCMC Pipeline}\label{ssec:mcmc}

For our MCMC sampler, we use a parallel-tempered \cite{2005PCCP....7.3910E} MCMC sampler enhanced by differential evolution \cite{original_de_storn} built in Python, similar to the sampler used in e.g. \cite{Cornish:2020vtw}. We include an array of different jump types, including prior draws and Fisher matrix proposals in a mixture of different parameter subspaces.  Including a diversity of different proposals helps prevent the sampler from getting stuck on secondary modes or collapsing into hyper-planes \cite{roberts_rosenthal_2007}.  

For the parallel-tempering temperature ladder, we can use either a standard geometric ladder $T_n = 10^{n\Delta \log T}$, or a more tuned ladder following the constant-entropy increase heuristic described in \cite{doi:10.1063/1.2907846}, such that
\begin{equation}\label{eq:ladder_entropy}
\int_{T_n}^{T_{n+1}}\frac{C_v(T)}{T}dT=\frac{1}{n_T}\int_{T=1}^{T=T_\text{max}} \frac{C_v(T)}{T}dT
\end{equation}
is a constant, where $C_v(T)$ is the heat capacity which can be estimated using:
\begin{equation}\label{eq:heat_cap}
C_v(T) = -\frac{\partial \langle\log L\rangle(T)}{\partial T}.
\end{equation}

While the heat capacity could be estimated adaptively during the burn-in, we currently use a short pilot run with a geometric ladder to estimate $\langle\log L\rangle(T)$ to obtain a useful temperature ladder for longer runs. Note that the temperature ladder described in Eq.~\eqref{eq:ladder_entropy} should make the exchange rates between adjacent-temperature chains approximately constant \cite{entropy_swap_acceptance}. We find that in practice the temperature ladder described in Eq.~\eqref{eq:ladder_entropy} typically achieves more efficient mixing, even if our estimator of $\langle\log L\rangle(T)$ is relatively poorly converged. We leave adaptive techniques to determine an optimal temperature ladder during burn-in to future work.  

We find that using $n_\text{chain}>100$  parallel tempering chains ensures robust convergence and mixing. We typically add an additional $\sim10-30\%$ of the chains set to $T=1$, in order to improve the sampling efficiency for the $T=1$ posterior. We also anchor the temperature ladder at high temperature with a single $T=\infty$ chain. Because we use a high density of chains, proposing exchanges only between adjacent temperatures can result in undesirably high exchange acceptance rates. Therefore, we instead make exchange proposals between uniformly random pairs of temperatures. An exchange proposal between a pair of temperatures $\{T_1,T_2\}$ with log-likelihoods $\{\log \mathcal{L}_{1}, \log \mathcal{L}_{2}\}$ respectively is accepted if

\begin{equation}\label{eq:exchange_acceptance}
\frac{1}{T_1}\left(\log \mathcal{L}_{2}-\log \mathcal{L}_{1}\right)+\frac{1}{T_2}\left(\log \mathcal{L}_{1}-\log \mathcal{L}_{2}\right)>\log p,
\end{equation}
where $p\in (0,1)$ is a uniform random number.

The non-exchange proposals for a chain at temperature T to advance from the state $\vec{\theta}_1$ to a state $\vec{\theta}_2$ are accepted if
\begin{equation}\label{eq:proposal_acceptance}
\frac{1}{T}\left(\log \mathcal{L}_2-\log \mathcal{L}_1\right)+\log P_2-\log P_1+\mathcal{J}_2-\mathcal{J}_1+H_{12}>\log p,
\end{equation}
where $p\in (0,1)$ is a uniform random number, $\log \mathcal{L}$ is the log-likelihood, $\log P$ is the prior, $H_{12}$ is the Hastings ratio for the proposal, and $J$ is a Jacobian factor that converts the prior on the mass parameters the chain actually uses, $\mathcal{M}_c$ and $M_t$, to a prior on $m_1$ and $m_2$. 

The Jacobian factor is computed according to:

\begin{equation}\label{eq:jacobian}
\mathcal{J}=\log\left(\frac{M_t^2\eta}{\sqrt{1-4\eta}}\right).
\end{equation}

Note that while the total mass $M_t$ is not dimensionless, the units cancel when calculating $J_2-J_1$ and can therefore be disregarded for this calculation.

For the prior factor, we use priors motivated by the most recent available LIGO/Virgo data release \cite{LIGOScientific:2021psn,Callister:2021fpo,Roulet:2021hcu,Biscoveanu:2022qac},
\begin{equation}\label{eq:prior}
\log P = -0.095 \frac{m_1}{m_\odot}+6.4q-\frac{(\chi_+-\mu_{\chi_+})^2}{2\sigma_{\chi_+}^2}-\frac{\chi_-^2}{2\sigma_{\chi_-}^2},
\end{equation}
where $\sigma_{\chi_+} =\sigma_{\chi_-} = 0.145/\sqrt{2}$, $\mu_{\chi_+}=0.06$, and $q=m_2/m_1$ is the ordinary mass ratio. Additionally, we enforce $0.2<q\leq1$, $m_1>5 m_\odot$, as well as the physical constraints $-1\leq \chi_1\leq 1$, and $-1\leq \chi_2\leq 1$.  These priors center the spins around zero, and give the sampler a preference for equal mass ratios and relatively low masses. The position, polarization, and phase parameters $\phi$, $\cos\theta$, $\cos i$, $\psi$, $\phi_c$ are only restricted to their physical ranges. 

For the purposes of this paper, we want to isolate the effect of a single-source SOBHB parameter estimation pipeline. The initial search pipeline and assessing the significance of candidate sources are separate problems \cite{2009PhRvD..80f3007L}, left to future work beyond the scope of this paper.
In order to represent the assumed effect of the search pipeline, we impose additional artificial hard boundaries on the two parameters which primarily control the shape of the time-frequency track: $f_i\in[f_{i,\text{true}}-\Delta f,f_{i,\text{true}}+\Delta f]$ and $\mathcal{M}_c\in[\mathcal{M}_\text{true}-\Delta \mathcal{M}_c,\mathcal{M}_\text{true}+\Delta \mathcal{M}_c]$. The hard lower bound on $\mathcal{M}_c$ indirectly imposes a corresponding hard lower bound on $M_t$ due to the requirement that the mass ratio be physical. For the test cases presented in this paper we chose these ranges by hand to be very broad such that the hard boundaries have little to no impact on the shape of the resulting contours in any other parameters. Tighter starting parameter ranges result in more efficient burn-in, at the expense of a risk of missing some probability in distant secondary modes.  

We also restrict the distance to a limited range $\log D_L\in[\log D_{L,\text{min}},\log D_{L,\text{max}}]$. Unlike the broad artificial ranges we impose on $\mathcal{M}_c$ and $f_i$, imposing an artificial maximum distance $\log D_{L,\text{max}}$ \emph{does} meaningfully impact the shape of the resulting parameter contours, at least for marginally significant sources in the presence of noise. This sensitivity occurs because for marginal sources there is potentially a large volume \cite{Moore:2019pke} of extremely faint distant candidate sources (with $S/N\sim 0$) with largely unconstrained parameters where noise fluctuations happen to produce log-likelihoods in the low significance tail of the posterior. This maximum distance cutoff is justified for the purposes of the parameter estimation pipeline presented in this paper, because assessing the significance of marginal candidate sources is a separate problem we leave to future work. Conversely, the minimum distance cutoff has no effect at all on the parameter estimation, and is included only to increase the acceptance of prior draw proposals, because noise fluctuations would not produce low-significance candidate sources which are substantially \emph{brighter} than the injected source.

The Hastings ratio $H_{12}$ is the difference between the proposal densities between the forward and reversed versions of the particular proposal type selected. In the current implementation, for SOBHBs the only proposals we have which require a density factor are the subset of prior proposals which include a Gaussian prior draw $\chi_s=\mathcal{N}(\mu_{\chi_{s}},\sigma_{\chi_{s}})$, $\chi_a=\mathcal{N}(0,\sigma_{\chi_{a}})$ in the spin parameters, for which 
\begin{equation}\label{eq:H12_chi}
H_{12}=\frac{(\chi_{+,2}-\mu_{\chi_{+}})^2-(\chi_{+,1}-\mu_{\chi_{+}})^2}{2\sigma_{\chi_{+}}^2}+\frac{\chi_{-,2}^2-\chi_{-,1}^2}{2\sigma_{\chi_-}^2}
\end{equation}
 is simply the factor which cancels the corresponding spin term in Eq.~\eqref{eq:prior}. For all other currently implemented proposal types, $H_{12}=0$.

 \subsection{Wavelet Coefficients}\label{ssec:wavelet_coeffs}

In order to perform likelihood evaluations, we need to first compute the wavelet domain representation of the proposed sources. For all the results in this paper, we use the frequency domain Taylor expansion method as described in \cite{2020PhRvD.102l4038C} to approximate the waveforms. For this method, we Taylor expand the waveform phase $\Theta(f)$ and amplitude $\mathcal{A}(f)$:

\begin{equation}\label{eq:phase_expand}
\begin{split}
\Theta(f)=\Theta(f_m)+2\pi(f-f_m)t(f_m)\\
+\pi(f-f_m)^2t'(f_m)+...
\end{split}
\end{equation}
\begin{equation}\label{amp_expand}
\mathcal{A}(f)=\mathcal{A}(f_m)+(f-f_m)\mathcal{A}'(f_m)+...
\end{equation}

where $f_m=m/(2 dt N_f)$ is the frequency of the $m$th frequency pixel. We then approximate the wavelet coefficients:
\begin{equation}
\begin{split}
w_{nm}=\mathcal{A}(f_m)(c_{nm}(t(f_m),t'(f_m))\cos\Theta(f_m)\\
-s_{nm}(t(f_m),t'(f_m))\sin\Theta(f_m)),
\end{split}
\end{equation}
where $c_{nm}$ and $s_{nm}$ are extracted from two pre-computed lookup tables
\begin{equation}
c_{nm}(t,t')=\int df \tilde{g}_{nm}(f)\cos(2\pi(f-f_m)t+\pi(f-f_m)^2t'),
\end{equation}
and $s_{nm}$ is identical except with the cosine replaced with sine. The WDM wavelets $\tilde{g}_{nm}$ are as described in \cite{2020PhRvD.102l4038C}. The amplitudes and phases for each of the A,E, and T channels are evaluated using the LISA response as described in Appendix B of \cite{Cornish:2020vtw} (see also \cite{Adams:2010vc,Rubbo:2003ap}). The wavelet coefficients $w_{nm}$ for the candidate signal need only be generated sparsely, for the set of $n,m$ values which are within the bandwidth of the signal. 

The simulated LISA data can be obtained from either a direct wavelet transform of time or frequency domain datasets \footnote{Our Python implementation of the forward and inverse WDM wavelet transforms is available at \url{https://github.com/XGI-MSU/WDMWaveletTransforms}} or by generating a simulated dataset directly in the wavelet domain. 

\subsection{Likelihoods}\label{ssec:likelihoods}
Once we have the wavelet coefficients for both the data and the proposed source, it is straightforward to compute the log-likelihood:

\begin{equation}\label{eq:logL}
\log L = \sum \frac{2w^{AET}_{nm,\text{pred}}w^{AET}_{nm,\text{data}}-{(w^{AET}_{nm,\text{pred}})^2}}{2S^{AET}_{nm}}
\end{equation}
where the sum runs over the A, E and T channels, and the sparse set of $\{n,m\}$ values for which the predicted signal contains power. The ability to vary the power spectrum $S^{AET}_{nm}$ as a function of time, frequency, and channel while remaining diagonal is a substantial advantage of evaluating the likelihood in the wavelet domain.

\section{Results}\label{sec:results}

\subsection{GW150914-like Source}\label{ssec:gw150914}
Here, we evaluate the parameter estimation pipeline on a source with parameters generally comparable to the signal that would have been generated if the first detected gravitational-wave source, GW150914, had merged approximately eighteen months after the beginning of LISA data collection began. In particular, we chose $m_1=35\;m_{\odot}$, $m_2=30\;m_{\odot}$, $F_i=0.025\;\text{Hz}$, $D_l=0.4\;\text{Gpc}$, and $\cos i=0.85$ as parameters representative of a generally similar source. The spin, position, and phase parameters were arbitrarily chosen to be $\chi_1\simeq-0.0293$, $\chi_2\simeq0.0713$, $\cos \theta\simeq0.331$, $\phi\simeq6.044$, $\psi\simeq5.376$, and $\phi_c\simeq1.217$. The corresponding chirp total mass terms for the search pipeline are $\mathcal{M}_c\simeq28.19\;m_\odot$, $M_\text{tot}\simeq65\;m_\odot$. These parameters result in a predicted merger time $t_c\simeq 1.474 \;\text{yr}$ after the start of the observing period. With the particular noise model used, the expected signal-to-noise ratio of this source is $S/N\simeq22.9$.

\begin{figure*}
\includegraphics[width=2\columnwidth]{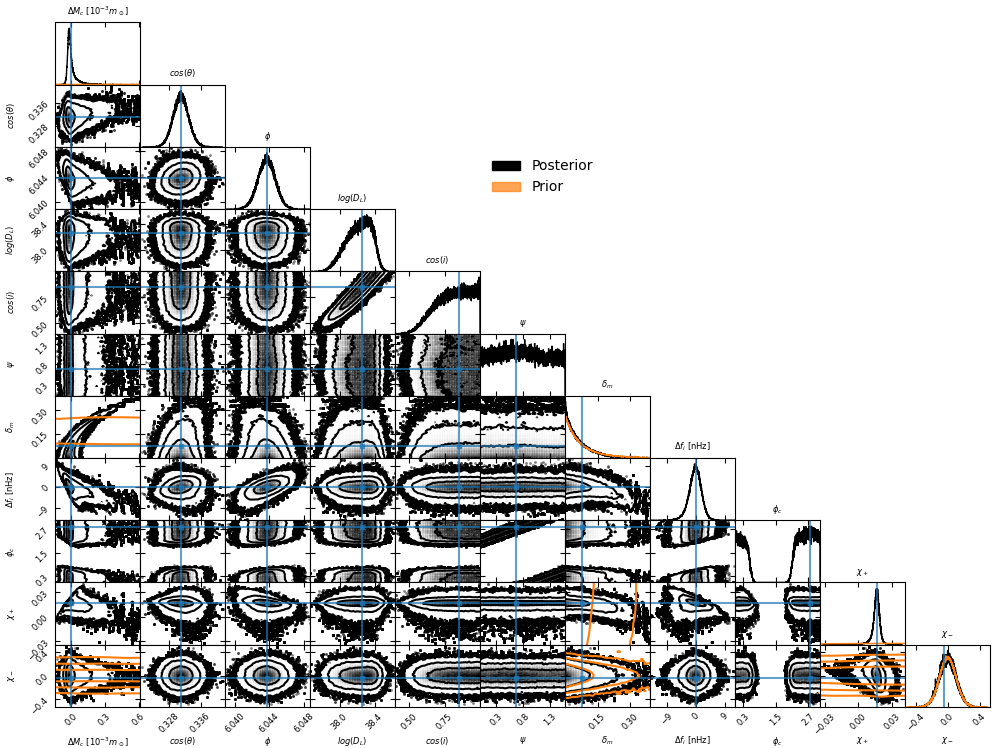}
\caption{\label{fig:2year_2015_ellipses} Corner plot of the parameter estimation pipeline results for a GW150914-like source that merges $t_c\simeq1.474\;\text{yr}$ into LISA's observation. Prior contours for combinations of $\mathcal{M}_c$, $\delta_m$, $\chi_+$, and $\chi_-$, the four parameters with non-uniform priors, are shown in orange. All parameters except $\chi_-$ and $\delta_m$ show constraints substantially stronger than their respective priors. Approximately $\sim50\%$ of the information contained in the posterior for $\delta_m$ originates from the prior, making it moderately prior informed. All other parameters are not meaningfully informed by their priors at all.}
\end{figure*}

\begin{figure}
\includegraphics[width=1\columnwidth]{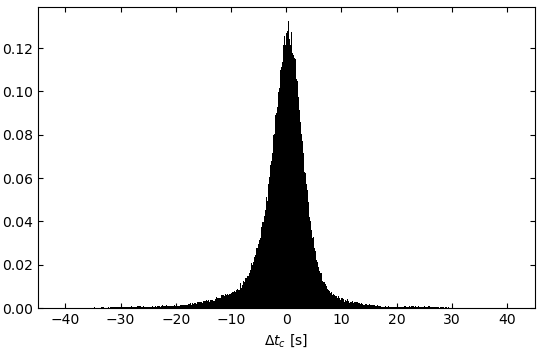}
\caption{\label{fig:2year_2015_tc} Forecast constraints on the merger time derived from the posterior in the parameter estimation run shown in Fig.~\ref{fig:2year_2015_ellipses}. The merger time, as would be observed by ground-based instruments, is forecast to within $\sigma_{t_c}\simeq 5.85\;s$. With our chosen sampling interval of $dt=1.8625\;s$, the source nominally continues to accrue non-zero LISA S/N until $\sim 23\;\text{hr}$ prior to merger. This final constraint is the constraint of interest for tests of GR. For the purpose of forecasting the timing of the merger in advance, $99\%$ of the LISA S/N has been accumulated by $\sim 18\;\text{days}$ prior to merger, so the merger time as observed by ground-based instruments would have been known to $\lesssim 10\;s$ $\sim$ by a month prior to merger, more than enough advance notice for any coordination efforts.}
\end{figure}

Results from running the parameter estimation pipeline on our GW150914-like source are shown in Fig.~\ref{fig:2year_2015_ellipses}. Almost all of the parameters we search over are significantly constrained compared to their respective priors. The resulting constraints on the forecast merger time, determined by a combination of $\mathcal{M}_c$, $M_t$, $F_i$, $\chi_+$, $\chi_-$, is shown in Fig.~\ref{fig:2year_2015_tc}. The merger time for this source can be predicted to within seconds, and the predicted merger phase $\phi_c$ is also reasonably well constrained.

The strength of the constraints depends strongly on how long LISA is able to observe the source for prior to merger. For example, for the same source with a merger after only $\sim6\;\text{months}$ of LISA observations, we would have $\sigma_{t_c}\simeq 52\;s$, compared to $\sigma_{t_c}\simeq 5.85\;s$ after $\sim18\;\text{months}$ of observations. Conversely, if the merger occurs after $\sim4\;\text{years}$ of observations, the constraints would have improved to $\sigma_{t_c}\simeq 2.48\;s$. All else being equal, mergers later in LISA's observation period will always produce better constraints. However, even for mergers quite early in the observation period, LISA will be able to produce useful forecasts. 

\subsection{Non-stationary Noise For a Yorsh-like Source}\label{ssec:yorsh}

For a second example of our parameter estimation pipeline, we consider a source with similar parameters as the SOBHB source in the upcoming Yorsh data release from the LISA data challenge (LDC) working group\footnote{\url{https://lisa.pages.in2p3.fr/LDC/data_generation/installation.html}} \cite{Baghi:2022ucj}. The Yorsh dataset contains two full years of LISA data this source, ending while the source is still $\sim2.7\;\text{years}$ prior to merger. Therefore, the source represents an interesting test of the constraints that could be obtained far in advance of merger. Additionally, the injected masses of $m_1\simeq79.2\; m_\odot$, $m_2\simeq54.1 \;m_\odot$ are relatively far into the high-mass tail of the LIGO/Virgo-informed priors we use, such that the priors significantly bias parameter estimation towards lower masses, especially at lower $S/N$. 

Because this source does not merge during the observation window, we want to investigate how the presence of non-stationary noise background could bias parameter estimation. Therefore, we change several parameters from the Yorsh dataset; namely, we set $\cos \theta=0.95$, $\cos i=0.07$, and $D_l\simeq0.3\;\text{Gpc}$. We also use a different noise model from the Yorsh dataset, inject a simulated non-stationary noise component, and generate our simulated dataset at $dt=30\;s$ directly in the wavelet domain. The remaining parameters are: $F_i\simeq0.0105\;\text{Hz}$, $\chi_1\simeq-0.0293$, $\chi_2\simeq0.0713$, $\phi\simeq-0.8777$, $\psi\simeq0.203$, and $\phi_c\simeq1.217$. 

\begin{figure}
\includegraphics[width=1\columnwidth]{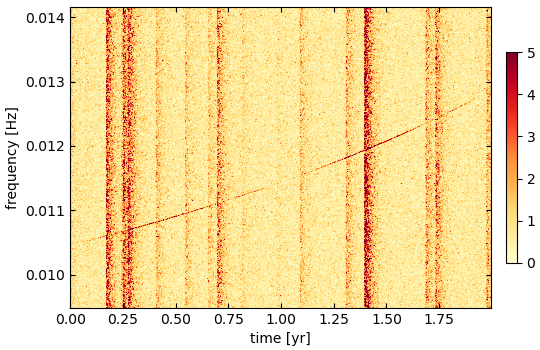}
\caption{\label{fig:power_blips} The whitened A-channel power of the injected non-stationary noise realization, along with the waveform track of the Yorsh-like source. To enable easier visualization of the source track, we have exaggerated the amplitude of the track by lowering the source distance to $D_l=0.03\;\text{Gpc}$. At $D_l=0.3\;\text{Gpc}$, as used in the test run of the pipeline, the track would not be visually discernable on this scale.}
\end{figure}

The injected non-stationary noise profile is shown in Fig.~\ref{fig:power_blips}. For this noise realization, we injected 20 `blips'; impulsive increases in the noise level in all three channels, with arbitrary amplitude and start times, which exponentially decay with an e-folding scale of $\sim 8.8\;\text{days}$. We fixed the total integrated power in the non-stationary component of the noise to be $50\%$ of the stationary component of the noise power in the same band. 

Although this type of noise does not necessarily represent any specific expected non-stationary type of noise, we chose it to reflect a physically plausible \cite{Edwards:2020tlp} kind of aperiodic non-stationary which could potentially arise from instrumental effects. For example, it could perhaps represent an unexpected ringdown period from micro-meteoroid impacts, or some type of spacecraft charging and discharging cycle from solar wind events. 

Subject to the $50\%$ power constraint, the blip amplitude and start times were chosen by hand, to produce an example of a noise realization that results in a relatively severe bias to parameter estimation if not modeled correctly. Most plausible non-stationary noise profiles would probably generate less bias than this particular combination of noise realization and source parameters. However, it should still be generally indicative of the kinds of parameter estimations biases that could arise in principle if non-stationarity is not modeled correctly, and the power of the wavelet-based noise modeling. 

\begin{figure*}
\includegraphics[width=2\columnwidth]{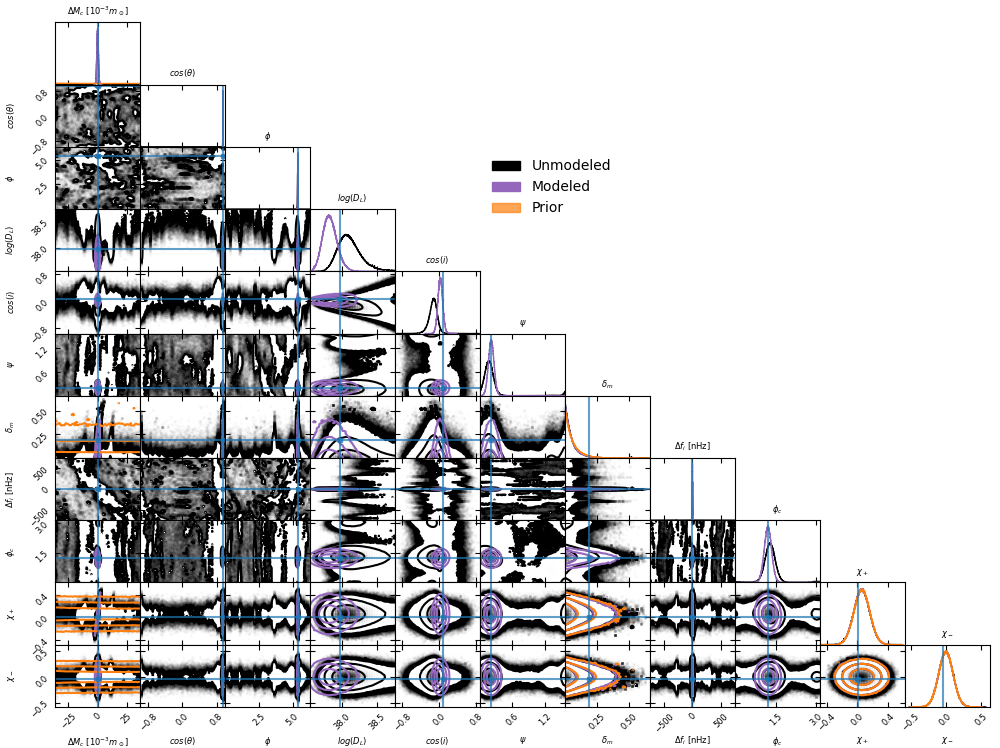}
\caption{\label{fig:2year_bias_ellipses} Corner plot of the parameter estimation pipeline results for a Yorsh-like source in the presence of the non-stationary noise in Fig.~\ref{fig:power_blips}. The black points represent running the parameter estimation pipeline incorrectly using a stationary noise model for $S^{AET}_{nm}$ in Eq.~\eqref{eq:logL}, and the purple contours represent the results of running the pipeline using the known-correct non-stationary noise model. While the central region of some parameters like $\mathcal{M}_c$, $F_i$, and the position parameters $\cos\theta$ and $\phi$ are well constrained in either case, with the incorrect stationary noise model there is significant additional structure that is not present when the noise is modeled correctly, including marginal secondary peaks many standard deviations form the central mode and some new band structure. The posteriors spin parameters and total mass are significantly informed by the prior in either case. Because the LIGO-based priors favor significantly lower total mass than the true injected source, the priors cause a significant bias in the total mass, separate from any bias caused by the non-stationary noise.}
\end{figure*}

To examine the bias that can be introduced by incorrectly modeling non-stationarity of the type shown in Fig.~\ref{fig:power_blips}, we do two separate runs of our parameter estimation pipeline; one with the true non-stationary noise profile, and the second incorrectly assuming a stationary noise profile with the same noise power, leaving the variation unmodeled. Our wavelet-based method is inherently well-suited to substituting between the stationary and non-stationary noise models; this substitution would have been substantially more difficult in the time or frequency domains.  A comparison of the results from the two runs is shown in Fig.~\ref{fig:2year_bias_ellipses}. Ignoring the non-stationarity introduces substantial secondary modes, and biases the inclination and distance. Due to the bias in the inclination and distance, the best fit source is at a substantially lower S/N; in the incorrect stationary noise model, the posterior points have a median $S/N\simeq7.72$, whereas in the correct noise model, the posterior points have a median $S/N\simeq10.4$. 

In fact, due to the lower S/N, Fig.~\ref{fig:2year_bias_ellipses} somewhat understates the severity of the bias. Because the median S/N of the posterior points in the incorrect noise model is so low, the upper prior limit on the distance substantially truncates the tail of the distribution, as can be seen by the fact that the contours in the $log(D_L)-\cos i$ plane are visibly cut off at the right edge of the distribution. These more distant sources essentially represent ultra-low S/N source candidates which could fit the noise. As discussed in Sec.~\ref{ssec:mcmc}, this cutoff is justified because for the purposes of this paper we are interested in isolating the effect of the parameter estimation itself, rather than assessing the overall significance of the source candidate. However, it is interesting to note that the bias caused by mis-modeling in this case is so severe that it could potentially cause a full pipeline to fail to identify this source as significant at all. 

A second interesting note is that the priors we have chosen on $M_t$ favor much lower total masses than the truth in either case, which biases the posteriors. However, these priors are informed based on fits to real LIGO/Virgo data; introducing a bias is simply the function of Bayesian priors. If, by the time LISA launches, more prior support exists for such high-mass sources, the priors could be revised. If not, then biasing the parameter estimation for a known source class towards more `realistic' masses based on a known mass distribution is simply the `correct' function of Bayesian inference. 

\begin{figure}
\includegraphics[width=1\columnwidth]{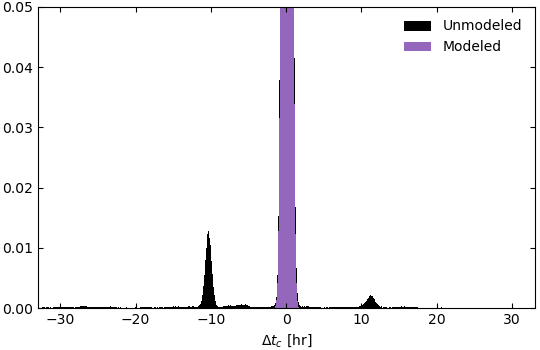}
\caption{\label{fig:2year_bias_tc} Forecast constraints on the merger time derived from the posterior in the parameter estimation run shown in Fig.~\ref{fig:2year_bias_ellipses}, with and without proper modeling of the non-stationary noise. Because the source is still $\sim2.7\;\text{yr}$ away from merger by the end of the two year observation window, the constraints on the merger time are much weaker than those shown in Fig.~\ref{fig:2year_2015_tc}, which extended much later into the inspiral. The constraints are also significantly degraded by the mis-modeling. With the correct model, $\sigma_{t_c}\simeq0.4\text{hrs}$, whereas with the incorrect stationary noise model $\sigma_{t_c}\simeq2.5\text{hrs}$. Additionally, the incorrect model produces distant secondary modes separated by over 10 hours from the truth.   }
\end{figure}

\begin{figure}
\includegraphics[width=1\columnwidth]{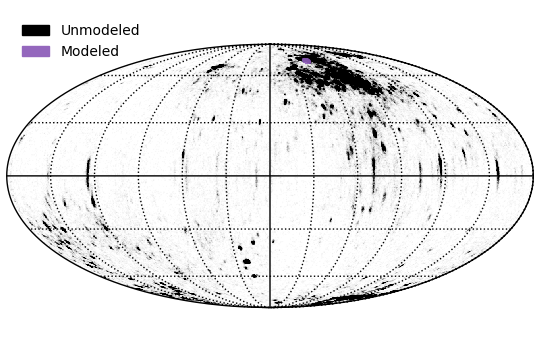}
\caption{\label{fig:skymap_bias} Skymap in ecliptic coordinates for the parameter estimation run in Fig.~\ref{fig:2year_bias_ellipses}. The black region is saturated at approximately the density of the 99.5\% probability contour. The dominant mode in both cases is at the correct sky location near the north ecliptic pole. The run with the non-stationary noise left unmodeled also exhibits a substantial secondary mode near the southern ecliptic pole that is contained within its 95\% probability contour, as well as a variety of other less significant mode structure. }
\end{figure}

The forecast constraints on the merger time are shown in Fig.~\ref{fig:2year_bias_tc}. Although the constraints are weaker than in Fig.~\ref{fig:2year_2015_tc}, with correct noise modeling LISA is still able to forecast the merger time as observed by ground-based observatories for an $S/N\sim{10.4}$ source to within better than half an hour 2.7 years in advance, which would give more than ample time for any reasonable coordination efforts, including potentially changing the configuration of ground-based gravitational-wave detectors to better optimize the multi-wavelength science yield. 

The forecast skymap for both runs is shown in Fig.~\ref{fig:skymap_bias}. The version with the biased noise model exhibits significant secondary structure; the 95\% probability contour covers $\sim3\;\text{deg}^2$, while the 99.7\% probability contour covers $\sim4100\;\text{deg}^2$. The 95\% probability contour includes a significant secondary mode near the southern ecliptic pole, which is not present at all in the posterior for the corrected noise model. The version with the corrected noise model is constrained to a tight ellipse around the true sky position, with a 95\% probability contour with area $\sim1.2\;\text{deg}^2$, and a 99.7\% probability contour of $\sim2.4\;\text{deg}^2$. The sky location predictions are of particular concern for potential electromagnetic follow-up searches, where biased sky maps could potentially lead observers to search for counterparts in the wrong places, or simply conclude the source is not well-localized enough to be an interesting counterpart search target. Therefore, our method's improved ability to account for non-stationary noise can substanially improve the utility of our results to multi-messenger observing partners. 

\subsection{Efficiency}\label{ssec:efficiency}

For any production MCMC pipeline, efficiency is crucial. Searching over a large parameter space and obtaining detailed contours requires large numbers of likelihood evaluations, and becomes limited by the expense of compute time. Because the SOBHB search will be just one component of a global fit, the efficiency of the entire pipeline can be limited if any individual part is slow. Efficiency will be of particular importance for online forecasting pipelines intended to facilitate coordination with multi-messenger and multi-wavelength observing partners,  which much report results as quickly as possible to maximise coordination. 

Even for offline analysis, a more efficient pipeline can be run for longer and obtain better converged results for a fixed amount of available compute time, or save financial and computational cost by running for less overall time. If likelihood evaluations are too slow, it may become impractical to obtain well converged results. 

For testing, we evaluate timings on a single core of an AMD Ryzen Threadripper 3970X. Our parallel tempering MCMC pipeline is parallelized at the level of the different parallel tempering chains, such that multiple cores allows us to do more likelihood evaluations in approximately the same amount of time, rather than performing individual likelihood evaluations faster. Therefore testing on a single core best isolates the computational efficiency of the likelihood evaluation itself.

For comparison purposes, we also estimate the time that a frequency-domain based pipeline would take to perform the same operations. This is possible because the evaluation of the waveform is essentially the same, and the time to calculate likelihoods can be estimated assuming a diagonal noise covariance matrix. This is only done for comparison purposes as the pipeline is not built to search in the frequency domain, so the timings should be considered approximate. 

The limits on computational efficiency are dictated by the speed of likelihood evaluations. Therefore, we want to isolate the speed of individual likelihood calculations. For the GW150914-like source, the track of the source is described by 65,445 time-frequency pixels in each of the three channels in our frequency-domain Taylor approximation based approach, while in the frequency domain the source is described by 30,437,374 real parameters per channel. Therefore, in this case the wavelet domain method achieves an inherent $\sim465\times$ compression. 

In this case, the average time per likelihood evaluation for our pipeline is approximately $8.9\;\text{ms}$. Our comparison frequency-domain based likelihood evaluation takes approximately $14s$, which is approximately $1600\times$ slower, far too slow to be practical for a realistic search pipeline. The advantage of the wavelet-based pipeline gets more significant as the observation time increases, or as the source sweeps through a larger frequency band. 

%\begin{center}
\begin{table*}
\begin{tabular}{||c c c c c c c c c c c||} 
 \hline
 source & $T_\text{obs} [\text{yr}]$ & $t_c [\text{yr}]$& $dt [\text{s}]$ & $N_f$ &$n_\text{WDM}$ & $n_\text{f}$ & $n_\text{f}/n_\text{WDM}$ & $t_\text{WDM} [\text{ms}]$& $t_\text{f} [\text{ms}]$ & $t_\text{f}/t_\text{WDM}$  \\ [0.5ex] 
 \hline\hline
 %0.5 & 6987 & 6893 & 826 & ... & ...\\ 
 %\hline
  GW150914 &0.5  &0.497& 1.875 & 4,096 &57,495 & 7,227,488& 126 & 3,500 & 5.6 & 600\\
  \hline
 GW150914 & 2.0  &1.47& 1.875 & 4,096 & 65,445  & 30,437,374 & 465 & 14,000 & 8.9 & 1600 \\
 \hline
 GW150914 & 4.0  &3.87& 1.875 & 4,096 &77,224  & 62,765,886 & 813 & 33,000 & 15 & 2300 \\
 \hline
  Yorsh-like & 2.0 &4.69& 30.0& 1,024 &4,414  & 308,750 & 70 & 180 & 0.52 & 350 \\
 \hline
 Yorsh-like & 2.0 &4.69& 30.0& 2,048 &5,760  & 308,750 & 54 & 180 & 0.63 & 290 \\
 \hline
  Yorsh-like & 2.0 &4.69& 30.0& 4,096 &9,975 & 308,750 & 31 & 180 & 0.98 & 180 \\
 \hline
 Yorsh-like & 5.32 &4.69& 10.0&2,048 &32,996 & 13,274,960 & 402 & 5,900 & 5.3& 1,100 \\
\hline
Yorsh-like & 5.32 &4.69& 5.0&4,096 &65,821 & 30,052,176 & 457 & 14,000 & 10& 1,400 \\
\hline
Yorsh-like &4.79 &4.69& 1.0& 16,384& 265,618 & 147,842,916 & 557 & 81,000 & 41& 2,000 \\
 \hline
 \hline
\end{tabular}
\caption{\label{tab:eff}Comparison of computation times and compression efficiencies between wavelet and frequency domains for different observation durations, sampling frequencies, and wavelet grid shapes. The wavelet domain improvement is always orders of magnitude, and is larger when the source is observed for longer or extends across a larger range of frequencies.}
\end{table*}
%\end{center}

Some timing results for various merger times, observation times, sampling frequencies, and wavelet parameters are shown in Table~\ref{tab:eff}. The wavelet-domain-based method is always at least two orders of magnitude for the sources, as expected given the level of compression of the track. For the two year Yorsh-like dataset, the wavelet domain likelihood calculations consistently take less than a millisecond. Assuming somewhat pessimistically that the sampler is able to generate on average one effective sample per 1,000 likelihood evaluations, then our pipeline should be able to generate 10,000 effective samples from the Yorsh-like likelihood in approximately ten minutes on a 16-core machine. With this effective sample generation rate, we were able to generate the well-converged 99.7\% probability contours for the complicated likelihood shown in Fig.~\ref{fig:2year_bias_ellipses} in less than a day on 16 cores, whereas obtaining a similar level of convergence with a frequency domain pipeline would have taken approximately a year. For most realistic production applications, generating well-converged 99.7\% probability contours will not be necessary, and the sampler could obtain adequate parameter estimation results in a few minutes. 

\section{Conclusions}\label{sec:conclusion}

In this paper, we have developed a wavelet-based parameter estimation pipeline for stellar-origin black hole binaries in LISA data. We have showcased this pipeline for two sources; a GW150914-like source, and a higher-mass source several years from merger in the presence of aperiodic non-stationary noise. The pipeline is able to efficiently characterize such SOBHB sources, and forecast merger times years in advance, enabling a variety of types of coordination with ground-based observatories, and potentially unlocking powerful tests of general relativity.   

We have shown that a wavelet-based approach is well-suited for analysis of SOBHB sources, due to the inherent data compression in the wavelet domain. In addition, it is well-suited for modeling types of astrophysical or instrumental non-stationarity that might plausibly occur in LISA data. Such modeling of non-stationarity is inherently more difficult in time or frequency-domain-based analyses, and we have shown that failing to correctly model non-stationarity can induce significant bias in parameter estimation, or potentially cause moderate S/N sources to be missed entirely. 

Failing to correctly subtract moderate S/N sources could also have cascading effects, making the entire global fit enterprise more difficult and less reliable. Therefore, it is essential for data analysis pipeline development efforts to ensure that they incorporate sufficient flexibility to handle non-stationary noise. 

In future work, we plan to extend our methodology to more source classes. Such source classes could include sources with detectable eccentricity, spin precession, and larger mass ratios. Additionally, our methodology can be incorporated into overall global fit efforts. 
%%%%%%%%%%%%%%%%%%%%%%%%%%%%%%%%%%%%%%%%%%%
%%%%%%%%%%%%%%%%%%%%%%%%%%%%%%%%%%%%%%%%%%%

\acknowledgements
We thank Will Farr for helpful input on the priors, and the LISA Data Challenge working group for providing us with access to the Yorsh dataset. This work was supported by NASA LISA Preparatory Science Grant 80NSSC19K0320.

%%%%%%%%%%%%%%%%%%%%%%%%%%%%%%%%%%%%%%%%%%%
%%%%%%%%%%%%%%%%%%%%%%%%%%%%%%%%%%%%%%%%%%%

\appendix

%%%%%%%%%%%%%%%%%%%%%%%%%%%%%%%%%%%%%%%%%%%
%%%%%%%%%%%%%%%%%%%%%%%%%%%%%%%%%%%%%%%%%%%

\bibliographystyle{JHEP}
\bibliography{sobhb_bib}
\end{document}